\documentclass[%
 reprint,
 superscriptaddress,
 amsmath,amssymb,
 aps,
]{revtex4-2}

\usepackage{graphicx}
\usepackage{dcolumn}
\usepackage{bm}
\usepackage{tabularx}

\usepackage{enumitem} 

\usepackage{color}
\usepackage{url}

\usepackage{mleftright} 

\usepackage[colorlinks]{hyperref}
\hypersetup{%
        plainpages=true,
        breaklinks=true,
        hypertexnames=false,
        pageanchor=true,
        colorlinks=true,
        linkcolor={blue},
        citecolor={magenta},
        urlcolor={blue},
        anchorcolor={black}
      }
\usepackage[all]{hypcap} 

\newcommand{\figref}[1]{\mbox{Fig.~\ref{#1}}}
\newcommand{\tabref}[1]{\mbox{Table~\ref{#1}}}
\newcommand{\secref}[1]{\mbox{Sec.~\ref{#1}}}

\renewcommand{\eqref}[1]{\mbox{Eq.~(\ref{#1})}}

\newcommand{\figurepanel}[2]{Fig.~\hyperref[#1]{\ref*{#1}(#2)}}
\newcommand{\figurepanelNoPrefix}[2]{\hyperref[#1]{\ref*{#1}(#2)}}

\newcommand{\ket}[1]{|#1\rangle}

\newcommand{\ketbra}[2]{\mleft| #1 \rangle \langle #2 \mright|}

\newcommand{\expec}[1]{\mleft\langle #1 \mright\rangle}

\newcommand{\nn}{\nonumber}

\newcommand{\be}{\begin{equation}}
\newcommand{\ee}{\end{equation}}
\newcommand{\bea}{\begin{eqnarray}}
\newcommand{\eea}{\end{eqnarray}}

\begin{document}


\title{Synthesizing electromagnetically induced transparency without a control field in waveguide QED using small and giant atoms}

\author{Andreas Ask}
\affiliation{
 Department of Microtechnology and Nanoscience (MC2), Chalmers University of Technology, SE-41296 Gothenburg, Sweden
}%

\author{Yao-Lung L. Fang}
\affiliation{
 Computational Science Initiative, Brookhaven National Laboratory, Upton, New York 11973, USA
}%

\author{Anton Frisk Kockum}
\affiliation{
 Department of Microtechnology and Nanoscience (MC2), Chalmers University of Technology, SE-41296 Gothenburg, Sweden
}

\date{\today}

\begin{abstract}

The absorption of photons in a three-level atom can be controlled and manipulated by applying a coherent drive at one of the atomic transitions. The situation where the absorption is fully canceled, and the atom thus has been turned completely transparent, has been coined electromagnetically induced transparency (EIT). The characteristics of EIT is a narrow transparency window associated with a fluorescence quench at its center frequency, indicating that inelastic scattering at this particular point is suppressed. The emergence of EIT-like transparency windows is common in waveguide quantum electrodynamics (QED) when multiple closely spaced quantum emitters are coupled to a waveguide. The transparency depends on the separation and energy detuning of the atoms. In this work, we study a number of different setups with \textit{two-level} atoms in waveguide QED that all exhibit EIT-like transparency windows. Unlike the case of a genuine \textit{three-level} atom, no drive fields are required in the systems we consider, and the coherent coupling of energy levels is mediated by the waveguide. We specifically distinguish between systems with genuine EIT-like dynamics and those that exhibit a transparency window but lack the fluorescence quench. The systems that we consider consist of both small and giant atoms, which can be experimentally realized with artificial atoms coupled to either photons or phonons. These systems can offer a simpler route to many EIT applications since the need for external driving is eliminated. 

\end{abstract}

\maketitle


\section{Introduction}
\label{sec:introduction}

The properties of atoms change when they interact strongly with an electromagnetic field. By shining light resonant with the atomic transitions, we can control and manipulate the quantum state of the atoms, and in the same process alter their spectroscopic behavior~\cite{CohenTannoudji1998}. In a three-level lambda ($\Lambda$) atom, for example, the absorption of photons can be canceled completely by having the atom interact with a strong, resonant control field. The process is enabled by the presence of a stable state in the $\Lambda$ system, allowing the atom to be driven into a completely dark state. This phenomenon has been named electromagnetically induced transparency (EIT)~\cite{Harris1989, Harris1999, Lukin2003, Fleischhauer2005}. The transparency comes with a steep dispersion relation, leading to a drastic change in the group velocity; as a result, EIT can be used to create ``slow'' light~\cite{Hau1999}, or even stop light completely~\cite{Hau2001}. Additionally, the intrinsic non-linearity offered by EIT has motivated a plethora of work in the optical domain for applications in quantum information processing~\cite{Lukin2000, Lukin2007, Fleischhauer2000, Beausoleil2004}.

In waveguide quantum electrodynamics (QED)~\cite{Gu2017, Roy2017}, atoms (or other quantum emitters) are made to interact with otherwise free photons propagating in a one-dimensional (1D) waveguide. The atoms could, for example, perform tasks as nodes in a larger quantum network~\cite{Kimble2008, Duan2010, Wehner2018}, or simulate quantum many-body physics~\cite{Bello2019, Wang2020, Mahmoodian2020}. Specifically, artificial atoms that are based on superconducting qubits and coupled to microwave transmission lines have seen rapid development in recent years~\cite{Gu2017, Astafiev2010, Hoi2011, Hoi2012, vanLoo2013, Koshino2013, Koshino2013a, Hoi2013, Hoi2013_2, Sathyamoorty2014, Inomata2014, Hoi2015, Forn-Diaz2017, Liu2017, Wen2018, Mirhosseini2018, Sundaresan2019, Wen2019, Mirhosseini2019, Lu2019, Kannan2020, Vadiraj2020}. Atoms made from superconducting circuits~\cite{You2011, Gu2017, Kockum2019a, Blais2020, Koch2007} are hard to use as $\Lambda$ systems directly, though, as their energy levels are usually formed in a ladder structure, lacking even a partially stable (or meta-stable) state. To observe EIT with artificial atoms, the level structure of the atom has to be purposefully engineered. This can be done either by embedding an artificial atom in a cavity (or resonator), forming an effective $\Lambda$ system in terms of dressed states, as done in Refs.~\cite{Koshino2013a, Inomata2014, Novikov2016, Long2018}, or to use giant atoms (atoms coupling to the waveguide at multiple points)~\cite{Kockum2014, Kockum2018, Kockum2020} to tune the relaxation rates of individual energy-levels to directly form a $\Lambda$ system. The latter approach was successfully demonstrated recently with artificial atoms coupled to a meandering transmission line~\cite{Vadiraj2020} and surface acoustic waves~\cite{Andersson2020}. 

\begin{figure*}[ht]
    \centering
    \includegraphics[width = \linewidth]{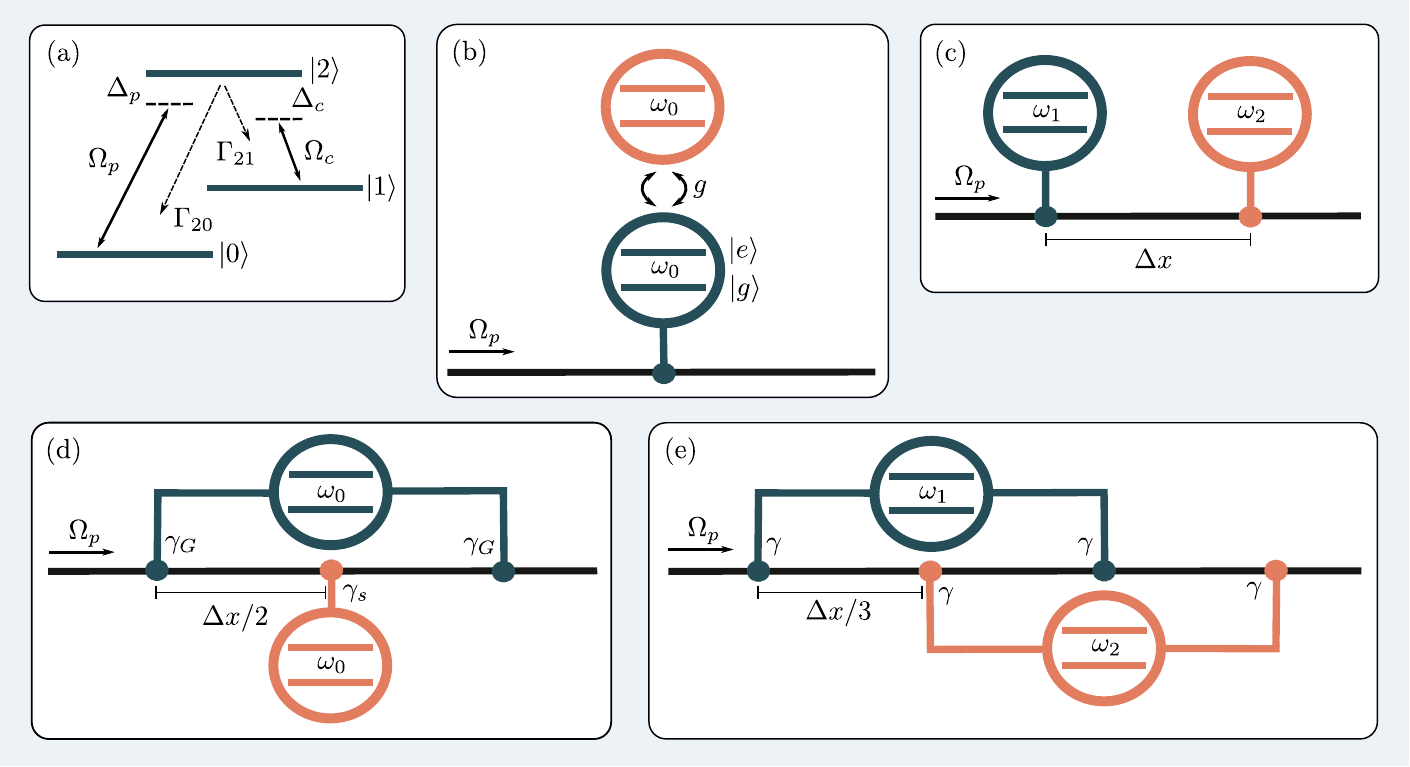}
    \caption{The setups considered in this work.
    (a) A three-level $\Lambda$ system with one transition driven by a control field $\Omega_c$ and the other by a weak probe field $\Omega_p$. The drives may be applied through a waveguide. We say that the system displays transparency if the probe field is perfectly transmitted.
    (b) A two-level system coupled to a waveguide and to another two-level system with dipole coupling $g$. The latter system does not interact directly with the waveguide.
    (c) Two two-level systems both coupled to a waveguide, separated by a distance $\Delta x$.
    (d) A giant atom coupled to the waveguide in two separate points, and a small atom coupled to the waveguide in-between those points.
    (e) Two giant atoms coupled to the waveguide in a braided configuration.
    \label{fig:setup}}
\end{figure*}
In this article, we take a different approach to EIT in waveguide QED. It is well known that multiple emitters in waveguides can form narrow transparency windows~\cite{Rephaeli2011, Fang2017, Mukhopadhyay2020}. It remains an open question, however, to what extent the transparency can be explained as a genuine EIT effect in some systems. A characteristic of EIT is the existence of a fluorescence quench~\cite{Zhou1996}. That is, no inelastic scattering takes place at the EIT frequency. This behaviour was reported for two co-located and non-identical atoms~\cite{Rephaeli2011, Fang2017}. Here, we consider multiple different systems in waveguide QED which all have narrow transparency windows even in the absence of a strong control field, see \figref{fig:setup}. In particular, we study two configurations with giant atoms. The extra coupling points of giant atoms lead to additional quantum interference effects, which have found several applications~\cite{Kockum2014, Gustafsson2014, Aref2016, Guo2017, Manenti2017, Kockum2018, Karg2019, Ask2019, Sletten2019, Gonzalez-Tudela2019, Andersson2020, Guimond2020, Bienfait2020, Guo2020, Kannan2020, Vadiraj2020, Wang2020, Zhao2020, Cilluffo2020, Kockum2020}, but the impact of these interference effects on transmission has not been thoroughly studied previously. The transparency windows we find in the systems in \figref{fig:setup} always stem from the existence of a dark state, but the origin of the dark state differs for the various systems considered. We show that the transparency only can be explained as a genuine EIT effect in some of these setups, as only those systems display a fluorescence quench.

Another feature of EIT is the existence of two fundamentally different regimes, in which the transparency has different origins. If the control field applied to a traditional $\Lambda$ system is strong enough, an Autler-Townes doublet~\cite{Autler1955} is formed, and the transparency is caused by scattering against two closely positioned, but distinctly different resonances. Alternatively, the transparency can be caused by quantum interference between two equal-energy states. These two regimes have been named the ATS and EIT regime, respectively~\cite{Abi-Salloum2010, Anisimov2011}. Experimentally, it can be hard to distinguish between these two regimes~\cite{Abdumalikov2010}, and a statistical measure was introduced as a way to aid in this distinction~\cite{Anisimov2011}. For the systems with multiple emitters in a waveguide that we consider, where no control field is applied, we show that the transparency can be caused by quantum interference, thus being in the ``EIT regime'', but yet miss a fluorescence quench; a potential source of confusion when it comes to classifying transparencies in waveguide QED as EIT-like or not. Our classification and explanations of which systems exhibit EIT-like transparencies can guide future work on applications of EIT-like behaviour in waveguide QED.

This article is organized as follows. In \secref{sec:method}, we use a single-photon scattering approach to derive a general set of equations that can be solved for the scattering coefficients for an arbitrary number of atoms in a waveguide. The atoms can be both small and giant, with coupling points arranged in any order. Additionally, we introduce the master equation, originally derived in Ref.~\cite{Kockum2018}, for multiple (possibly giant) atoms coherently driven from the waveguide. In \secref{sec:results}, we solve the system of equations derived from the scattering calculation for each setup in \figref{fig:setup}. From the transmission coefficients, we derive the equivalents of EIT and ATS regimes for each setup. We solve the master equation for each setup in the EIT regime and evaluate the existence of a fluorescence quench. We conclude in \secref{sec:conclusion}.


\section{Methods}
\label{sec:method}

To analyze the transmission properties of the setups in \figref{fig:setup}, we use two different methods:
\begin{enumerate}[label = (\roman*)]
    \item A single-photon-scattering calculation allows us to derive analytical expressions for the transmission coefficients. From the poles of the transmission coefficient we can distinguish between the EIT and ATS regimes, much in accordance with the analysis in \cite{Abi-Salloum2010} and \cite{Anisimov2011}.
    \item Once we have found the parameter threshold for the EIT regime, we numerically solve a master equation to study inelastic transmission properties.
\end{enumerate}


\subsection{Single-photon scattering}
\label{subsection:scattering}

The scattering approach we use was developed in Ref.~\cite{Zheng2013} and further illustrated in Refs.~\cite{Fang2014, Fang2015}. From this method, we derive a general set of equations that can be solved for the single-photon scattering amplitudes of systems consisting of an arbitrary number of small and giant atoms in a waveguide. The equations we derive are valid for all configurations of giant-atom coupling points, which can be both nested and braided~\cite{Kockum2018}, as well as surrounding one or multiple small or giant atoms.  

Starting from the Schr\"odinger equation (we set $\hbar$ and the wave velocity to $1$ throughout the paper) 
\begin{equation}
H \ket{\Psi(k)}_R = k \ket{\Psi(k)}_R,
\label{eq:shrodinger}
\end{equation}
where $\ket{\Psi(k)}_R$ denotes the scattering eigenstate of the waveguide and emitters with a right-going photon ($R$) injected at a far distance to the left. The total Hamiltonian has three parts:
\be
H = H_E + H_B + H_{\text{int}}.
\ee
We write the Hamiltonian describing the emitters as
\be
H_E = \sum_i \omega_i \sigma^+_i\sigma^-_i,
\ee
where $\omega_i$ is the emitter transition frequency and $\sigma^+_i (\sigma^-_i)$ is the creation (annihilation) operator for emitter $i$. The free-field Hamiltonian for the waveguide is given by
\begin{equation}
H_B = -i \int dx \left [ a_R^{\dagger}(x)\frac{d}{dx}a_R(x) - a_L^{\dagger}(x)\frac{d}{dx}a_L(x) \right],
\end{equation}
where $a^{\dagger}_\alpha(x)$ [$a_\alpha(x)$] creates [annihilates] a photon at position $x$ moving in the $\alpha = R, L$ direction. The interaction Hamiltonian can be written as two sums
\bea
H_\text{int} &=& \sum_{i=1}^{N} \sum_{j=1}^{M_i} \int dx\, \delta(x - x_{i_j}) \sqrt{\frac{\gamma_{i_j}}{2}} \nn\\
&&\times \mleft[ \mleft(a_R^{\dagger}(x) + a_L^{\dagger}(x) \mright) \sigma^-_i + \text{H.c.} \mright],
\eea
where $N$ is the number of emitters, $M_i$ is the number of coupling points for emitter $i$, $x_{i_j}$ is the position of the $j$th coupling point for emitter $i$, $\gamma_{i_j}$ is the corresponding coupling strength, and $\text{H.c.}$ denotes Hermitian conjugate.

By defining the state where both right and left-going modes in the waveguide are in the vacuum state and all emitters are in their ground state,
\be
\ket{0} \equiv \ket{0}_R \otimes \ket{0}_L \otimes \ket{g}_1 \otimes \cdots \otimes \ket{g}_N,
\ee
we can write the scattering eigenstate $\ket{\Psi(k)}_R$ as
\bea
\ket{\Psi(k)}_R &=& \int dx \mleft( \phi_R(k,x) a_R^\dag(x) + \phi_L(k,x) a_L^\dag(x) \mright) \ket{0} \nn\\
    && + \sum_{i=1}^N e_i(k) \sigma_i^+ \ket{0}.
\eea
The amplitudes $\phi_{R/L}$ contain a plane-wave ansatz
\begin{widetext}
\bea
  \phi_R(k,x) &=& \frac{e^{ikx}}{\sqrt{2\pi}} \mleft[ \theta(x_1 - x) + t_1 \theta(x - x_1) \theta(x_2 - x)
  + t_2 \theta(x - x_2) \theta(x_3 - x) + \ldots + t_{N'} \theta(x - x_{N'})) \mright], \\
    \phi_L(k,x) &=& \frac{e^{-ikx}}{\sqrt{2\pi}} \mleft[ r_1 \theta(x_1 - x) + r_2 \theta(x - x_1) \theta(x_2 - x) + \ldots + r_{N'} \theta(x - x_{N'-1}) \theta(x_{N'} - x) \mright],
\eea
\end{widetext}
where $r_{n}(k)$ and $t_n(k)$ are the reflection and transmission coefficient for the $n$th coupling point, respectively. Note that 
the running index $n\in[1, N']$ is simply counting from the first coupling point ($x_{1_1}$) at the very left to the last one at the very right
($x_{N'}$, which need not be $x_{N_{M_N}}$, since for giant atoms the last coupling point can belong to any of the atoms).

By plugging the plane-wave ansatz into \eqref{eq:shrodinger} and collecting coefficients for each basis state ($a_R^{\dagger}\ket{0}$, $a_L^{\dagger} \ket{0}$, and $\{\sigma_i^+\ket{0}\}$), it is possible to derive, after some manipulation~\cite{Zheng2013}, the following set of coupled equations for each atom $i$:
\bea
    e_i(k) &=& i\frac{e^{i k x_{i_j}}}{\sqrt{\pi \gamma_{i_j}}} \mleft( t_{i_j} - t_{i_j - 1} \mright), \label{eq:scattering1} \\
    e_i(k) &=& i \frac{e^{-i k x_{i_j}}}{\sqrt{\pi \gamma_{i_j}}} \mleft( r_{i_j} - r_{i_j + 1} \mright), \label{eq:scattering2}\\
    e_i(k) &=& \sum_{j=1}^{M_i} \frac{\sqrt{\gamma_{i_j}}}{2\sqrt{4\pi}(k - \omega_i)} \mleft[ e^{i k x_{i_j}} \mleft( t_{i_j - 1} + t_{i_j} \mright) \mright. \label{eq:scattering3} \\
    &&\mleft. + e^{-i k x_{i_j}} \mleft( r_{i_j} + r_{i_j + 1} \mright) \mright], \nn
\eea
where $t_{i_j}$ is defined as $t_n$ for $x_{i_j}$ being the $n$th coupling point, and similarly for $r_{i_j}$.  The first and last coupling points are exceptions for $t$ and $r$ respectively, where $t_0 = 1$ and $r_{N'+1} = 0$. Equations~(\ref{eq:scattering1}) and (\ref{eq:scattering2}) can be regarded as boundary conditions for $r$ and $t$, and \eqref{eq:scattering3} stems from energy conservation. Note that the sum of the overall transmission and reflection probabilities is conserved, $|t|^2 + |r|^2 = |t_{N'}(k)|^2 + |r_1(k)|^2 = 1$, if no loss or noise channel is present (which we assume). In total, this gives a system of $2 \sum_i M_i + N$ independent equations that can be solved for the scattering coefficients between adjacent coupling points. 


\subsection{Coherent drive}

The master equation and the corresponding input-output relations for an array of giant atoms with a coherent driving field in the waveguide was derived in Ref.~\cite{Kockum2018}. This master equation can be written, within the rotating-wave approximation (RWA), and in a frame rotating with the drive frequency $\omega_d$, as
\begin{widetext}
\begin{equation}
\begin{split}
    \dot{\rho} & = -i [H_{\text{driven}},\rho ] + \sum_{i=1}^N\sum_{j=1}^{M_i}\sum_{k=1}^{M_i} \sqrt{\gamma_{i_j}\gamma_{i_k}}\cos \phi_{i_ji_k}\mathcal{D}[ \sigma^-_{i} ] \rho + \sum_{i=1}^{N-1}\sum_{l=i+1}^N \sum_{j=1}^{M_i}\sum_{k=1}^{M_l} \sqrt{\gamma_{i_j}\gamma_{l_k}}\cos \phi_{i_jl_k} \\
    & \ \ \ \times \mleft[ \mleft( \sigma^-_i \rho \sigma^+_l - \frac{1}{2} \mleft\{ \sigma^+_i \sigma^-_l , \rho \mright\} \mright) + \text{H.c.} \mright],
    \label{eq:master}
    \end{split}
    \end{equation}
with
\begin{equation}
    \begin{split}
        H_{\text{driven}} & = \sum_{i=1}^N  \mleft( \Delta_i + \sum_{j=1}^{M_i - 1} \sum_{k=n+1}^{M_i} \sqrt{\gamma_{i_j}\gamma_{i_k}} \sin \phi_{i_ji_k} \mright) \sigma^+_i\sigma^-_i + \sum_{i=1}^{N-1}\sum_{l=i+1}^N \sum_{j=1}^{M_i}\sum_{k=1}^{M_l} \frac{\sqrt{\gamma_{i_j}\gamma_{l_k}}}{2}\sin \phi_{i_j l_k} \mleft( \sigma^-_i \sigma^+_l + \sigma^+_i \sigma^-_l \mright)  \\
        & \ \ \ \ - i \sum_{i=1}^N \sum_{j=1}^{M_i} \sqrt{\frac{\gamma_{i_j}}{2}} \mleft( \alpha e^{i\phi_{1_1 i_j}} \sigma^+_i - \text{H.c.} \mright),
    \end{split}
\end{equation}
\end{widetext}
where $|\alpha|^2$ is the number of photons per second coming from the coherent drive, 
$\phi_{i_jl_k}$ denotes
the phase shift acquired when moving from point $i_j$ to $l_k$, 
and the rest of the symbols ($N$, $M_i$, $\gamma_{i_j}$, etc.) have the same meaning as in the preceding subsection.
Using input-output theory, we find the output bosonic operator which has contributions from all of the atoms~\cite{Kockum2018},
\begin{align}
    b_{\text{out}} &= a_R\mleft(x\rightarrow (x_{N'})^+\mright)\\ 
    &= \alpha e^{i\phi_0} + \sum_{i=1}^N \sum_{j=1}^{M_i} e^{i\phi_{i_j N'}}\sqrt{\frac{\gamma_{i_j}}{2}}\sigma^-_i,
\end{align}
where $\phi_0$ refers to the total phase shift acquired by a photon traversing the entire system (from the first coupling point, $x_{1_1}$, to the last one, $x_{N'}$),
then the transmission coefficient, describing the transmission through the whole array of atoms, can be written as
\begin{equation}
    t = \frac{\langle b_\text{out} \rangle}{\alpha}
    = e^{i\phi_0} + \frac{1}{\alpha} \sum_{i=1}^N \sum_{j=1}^{M_i} e^{i\phi_{i_j N'}} \sqrt{\frac{\gamma_{i_j}}{2}} \langle \sigma^-_i\rangle.
    \label{eq:t}
\end{equation}
The first term in \eqref{eq:t} accounts for the propagation of the drive through the whole system, and the second term accounts for the contribution from each coupling point of each atom. 
In the weak-driving limit $\alpha\rightarrow0$, $t\rightarrow t_{N'}$ (up to an irrelevant global phase).

Equation~(\ref{eq:master}) was originally derived for an array of giant atoms, but can be used also for the cases where small atoms are placed inside giant atoms, by setting all but one coupling point to zero for the small atom, and adjust the phase shifts accordingly. 


\subsection{Inelastic scattering properties}

We study the inelastic scattering properties using the solution to the master equation of the coherently driven system, \eqref{eq:master}. To evaluate the existence of a fluorescence quench, we calculate the total inelastic photon flux~\cite{Zhou1996, Lalumiere2013, Fang2015}
\begin{equation}
    F(\omega_p) = \int d\omega S(\omega),
    \label{eq:flux}
\end{equation}
where $S$ is the inelastic power-spectrum (resonance fluorescence)
\begin{equation}
    S(\omega) = \int e^{-i\omega t} \expec{b_{\text{out}}^{\dagger}(t)b_{\text{out}}(0)}_{\rm ss}
\end{equation}
with $\langle\rangle_{\rm ss}$ denoting the expectation value in the steady state. Note that to access
the inelastic properties, the term containing a Dirac delta function at the drive frequency $\omega_p$, corresponding to elastic scattering, is dropped from the above definition.


\section{Results}
\label{sec:results}

In addition to a traditional three-level $\Lambda$ system [see \figurepanel{fig:setup}{a}] we study the following configurations of multiple emitters in a waveguide:
\begin{enumerate}[label = (\roman*)]
\item two dipole-coupled atoms, of which one is connected to the waveguide while the other atom is disconnected [see \figurepanel{fig:setup}{b}];
\item two atoms placed after each other in a waveguide, slightly detuned from each other [see \figurepanel{fig:setup}{c}];
\item a giant atom surrounding a small atom [see \figurepanel{fig:setup}{d}];
\item two giant atoms with coupling points in a braided configuration [see \figurepanel{fig:setup}{e}].
\end{enumerate}
All of these systems can be shown to exhibit transmission properties resembling EIT in some parameter regime, even without the application of a strong control field, since they are all capable of producing dark states. The main question we seek to answer is to what degree these systems and their transmission properties can be mapped to the three-level $\Lambda$ atom in \figurepanel{fig:setup}{a}. Here, an important factor is whether fluorescence quenches are exhibited at the EIT-like transmission peaks or not.


\subsection{Lambda system}

We first briefly review EIT in a traditional three-level $\Lambda$ system, to understand the role of the control field and the parameter regime in which EIT occurs. This allows us to compare the $\Lambda$ system to the other systems we consider later on. For a more detailed discussion of EIT, see, e.g., Refs.~\cite{Fleischhauer2005, Lukin2003}, and for a comparison of EIT in different three-level systems, see Ref.~\cite{Abi-Salloum2010}.

A $\Lambda$ system has the level structure seen in \figurepanel{fig:setup}{a}. Only the transitions $\ket{0} \leftrightarrow \ket{2}$ and $\ket{1} \leftrightarrow \ket{2}$ are allowed, as $\ket{1}$ is a dark state that does not couple to the environment. 
When a control field with amplitude $\Omega_c$ and frequency $\omega_c$ is applied at the $\ket{1} \leftrightarrow \ket{2}$ transition, and a probe field with amplitude $\Omega_p$ and frequency $\omega_p$ is applied at the $\ket{0} \leftrightarrow \ket{2}$ transition, the system dynamics are given by the master equation~\cite{Fleischhauer2005}
\bea
\dot{\rho} &=& -i [H,\rho] + \frac{\Gamma_{20}}{2} \mleft( 2\sigma_{02}\rho\sigma_{20} - \{\sigma_{22},\rho\} \mright) \nn\\
&&+ \frac{\Gamma_{21}}{2} \mleft( 2\sigma_{12}\rho \sigma_{21} - \{ \sigma_{22},\rho \} \mright) \nn\\
&&+ \gamma_{2\phi} \mleft( 2\sigma_{22}\rho \sigma_{22} - \{ \sigma_{22},\rho \} \mright) \nn\\
&&+ \gamma_{1\phi} \mleft( 2\sigma_{11}\rho \sigma_{11} - \{ \sigma_{11},\rho \} \mright),
\label{eq:master_lambda}
\eea
where $\sigma_{ij} = \ketbra{i}{j}$, $\Gamma_{ij}$ is the decay rate from state $\ket{i}$ to $\ket{j}$, and $\gamma_{i\phi}$ is the pure dephasing rate of state $\ket{i}$. The system Hamiltonian can be made time-independent by applying the RWA and going into a rotating frame:
\bea
H &=& \Delta_p \sigma_{22} + (\Delta_p - \Delta_c) \sigma_{11} + i\frac{\Omega_p}{2} \mleft( \sigma_{02} - \sigma_{20} \mright) \nn\\
&& + i\frac{\Omega_c}{2} \mleft( \sigma_{12} - \sigma_{21} \mright),
\label{eq:Hamil_lambda}
\eea
where $\Delta_p = \omega_2 - \omega_p$, $\Delta_c = (\omega_2 - \omega_1) - \omega_c$, and $\omega_i$ is the energy of state $\ket{i}$.

To observe EIT, we solve \eqref{eq:master_lambda} in the regime of a strong control field and a weak probe field. To make the analysis clearer, we assume the ideal case of no pure dephasing ($\gamma_{i\phi} = 0$) and perfect resonance of the control field ($\Delta_c = 0$). From the solution to the master equation and the input-output relation
\be
b_{\text{out}} = b_{\text{in}} + \sqrt{\frac{\Gamma_{20}}{2}}\sigma_{02},
\ee
we obtain the transmission coefficient  (up to the first order in the small parameter $\Omega_p/\Gamma_{20} \ll 1 $)
\begin{equation}
    t_{\Lambda} = 1 - \frac{2\Gamma_{20}\Delta_p}{4(\gamma_2 + i\Delta_p)\Delta_p - i\Omega_c^2},
    \label{eq:t_lambda}
\end{equation}
where $\gamma_2 = (\Gamma_{21} + \Gamma_{20})/2$. The EIT manifests itself as a transmission peak at $\Delta_p = 0$. The origin of the transmission peak can be seen by inspecting the complex roots of $t_{\Lambda}$~\cite{Abi-Salloum2010, Anisimov2011}
\begin{equation}
    \mathcal{Z_{\pm}} = \frac{i\gamma_2}{2} \pm \frac{1}{2}\sqrt{\Omega_c^2 - \gamma_2^2},
\end{equation}
which are purely imaginary for $\Omega_c < \gamma_2$. In this parameter regime, the high transmission is caused by quantum interference between two resonances positioned at the same energy. This is the EIT regime. For $\Omega_c > \gamma_2$, an Autler-Townes doublet is formed by the dressing of energy levels by the strong control field. This is referred to as the ATS regime. 

\begin{figure}[ht!]
    \centering
    \includegraphics[width=\columnwidth]{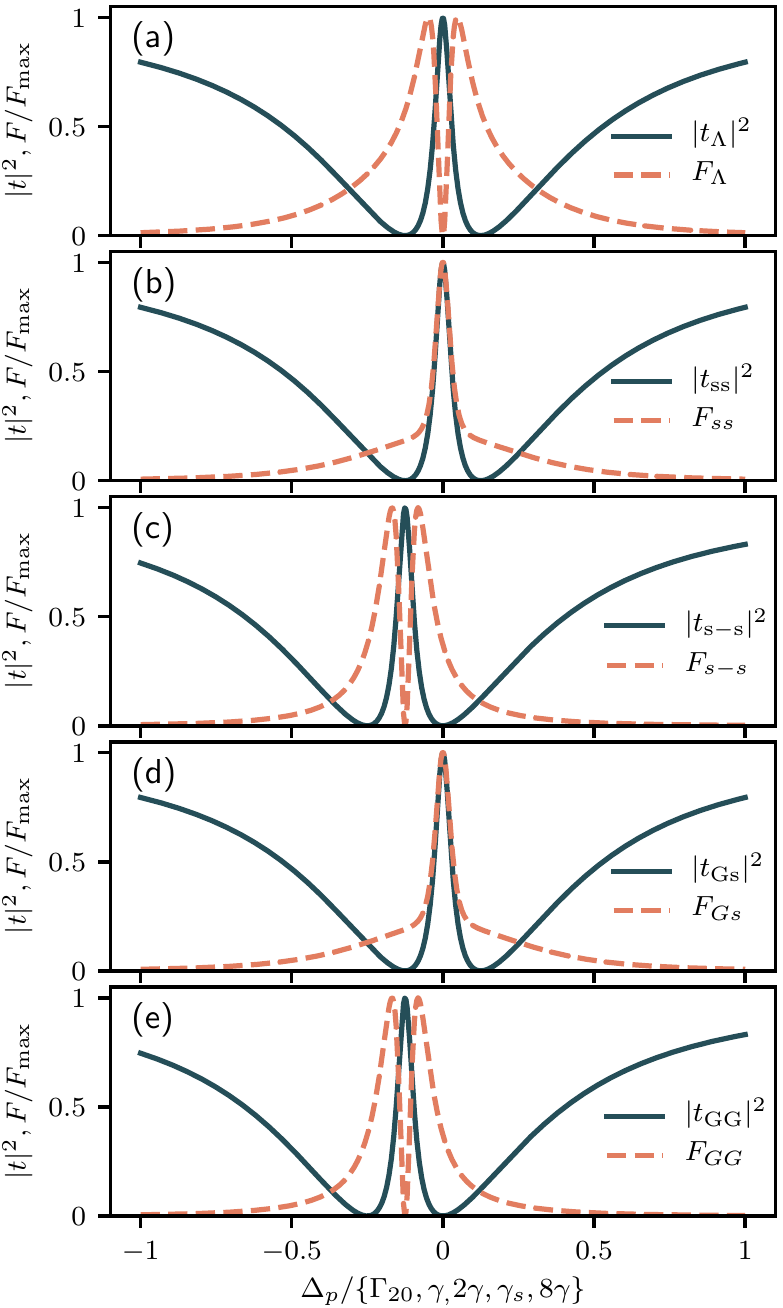}
    \caption{Single-photon transmission coefficient (solid lines) and inelastic photon flux (dashed lines) for the five systems depicted in \figref{fig:setup}.
    (a) A traditional $\Lambda$ system,
    (b) two dipole-coupled atoms with one atom disconnected from the waveguide,
    (c) two distant atoms with a small detuning,
    (d) a giant atom surrounding a small atom, where the giant atom is dark, and
    (e) two giant atoms, where both atoms are bright but detuned as in (c). All systems are in the parameter regime where they fulfil the EIT criteria, summarized for each system in \tabref{tab}. For the exact parameters used for each panel, see the main text. The EIT peak is only accompanied by a fluorescence quench when the atoms are detuned; see the dashed lines in (c) and (e). When a bright and a dark atom are coupled through a dipole coupling, either directly as in (b) or through the waveguide as in (d), the comparison with a $\Lambda$ system only holds in the single-photon regime, as is evident from the lack of fluorescence quench. We used a coherent drive amplitude of $|\alpha|^2 = \{\Gamma_{20}/10,\gamma/10,\gamma/5,\gamma_s/10,4\gamma/5$\} in (a)-(e), respectively, to calculate $F$.
    \label{fig:results}}
\end{figure}

We plot the transmission $t_{\Lambda}$ and the flux $F_{\Lambda}$ as a function of probe detuning $\Delta_p$ in the EIT regime in \figurepanel{fig:results}{a}. We use the parameters $\Gamma_{21} = 0$ and $\Omega_c = \Gamma_{20}/4$. The perfect transmission of probe photons at $\Delta_p = 0$ is due to destructive interference between the two different excitation pathways in the system, $\ket{0} \rightarrow \ket{2}$ and $\ket{0} \rightarrow \ket{2} \rightarrow \ket{1} \rightarrow \ket{2}$, which are equally likely in this parameter regime. The fluorescence is known to be quenched at the EIT peak in a $\Lambda$ system \cite{Zhou1996}, which manifests itself as zero flux at the EIT frequency. The reason for the quench can be understood in terms of dark states. In the EIT regime, there is zero probability of occupying the bright state $\ket{2}$, as the system is trapped in a dark superposition of $\ket{0}$ and $\ket{1}$. Thus, no photon absorption takes place at $\Delta_p = 0$ and the inelastic scattering is quenched. 

\begin{table}
   \caption{Criteria for being in the EIT regime for each system configuration, and whether the EIT peak is associated with a fluorescence quench or not.} 
   \label{tab:summary}
   \renewcommand{\arraystretch}{1.2}
   \renewcommand{\tabcolsep}{0.1cm}
   \centering 
  \begin{ruledtabular}
   \begin{tabular}{c | ccccc} 
     Setup & $\Lambda$ &  ss & s-s & Gs & GG\\ 
   \hline
     Schematic & \figurepanelNoPrefix{fig:setup}{a} & \figurepanelNoPrefix{fig:setup}{b} & \figurepanelNoPrefix{fig:setup}{c} & 
     \figurepanelNoPrefix{fig:setup}{d} & \figurepanelNoPrefix{fig:setup}{e} \\
     EIT regime & $\Omega_c < \gamma_2$ & $g < \gamma/4 $ & $\delta < \gamma $ & $\gamma_G < \gamma_s/16$ & $\delta < 4\gamma$  \\
     Quench & Yes & No & Yes & No & Yes \\
   \end{tabular}
   \end{ruledtabular}
   \label{tab}
\end{table}


\subsection{Two small atoms}

We consider two setups with small atoms. In the first setup, which can be seen in \figurepanel{fig:setup}{b}, only one of the atoms is connected to the waveguide, and the two (identical) atoms are dipole-coupled. We refer to this setup as ``ss''. In the second setup, we consider two detuned atoms coupled to the waveguide at separate points, see \figurepanel{fig:setup}{c}. The coupling between the atoms is mediated by the waveguide due to their separation in space (set to be one wavelength). We refer to this setup as ``s-s''.


\subsubsection{Two dipole-coupled small atoms with one detached from the waveguide (``ss'')}
\label{subsubsec:ss}

In this case, the system of equations given by Eqs.~(\ref{eq:scattering1})-(\ref{eq:scattering3}), which we derived in \secref{subsection:scattering}, needs to be modified since one of the atoms is decoupled from the waveguide. By adding the dipole coupling to the emitter Hamiltonian
\be
H_E = \sum_{i=1}^2 \omega_i \sigma_i^+\sigma_i^- + g \mleft( \sigma^+_1 \sigma^-_2 + \sigma^+_2\sigma^-_1 \mright),
\label{eq:H_E-ss}
\ee
it is possible to derive the modified set of equations
\begin{eqnarray}
e_1(k) &=& \frac{i}{ \sqrt{\pi \gamma}} (t - 1), \label{eq:1}\\
e_1(k) &=& \frac{i r}{\sqrt{\pi \gamma}},  \label{eq:2}\\
e_1(k) &=& \frac{\sqrt{\gamma}}{2\sqrt{4\pi}(k-\omega_1)} \left ( 1 + t \right ) + \frac{g e_2(k)}{k - \omega_1}, \label{eq:3} \\
e_2(k) &=& \frac{ge_1(k)}{(k-\omega_2)}, \label{eq:4}
\end{eqnarray}
where we position the waveguide-coupled atom at $x = 0$. We solve Eqs.~(\ref{eq:1})-(\ref{eq:4}) for $\omega_1 = \omega_2 \equiv \omega_0$, and obtain the single-photon transmission coefficient
\begin{equation}
t_{ss} = 1 - \frac{\gamma \Delta_p}{(\gamma - 2 i \Delta_p) \Delta_p + 2 i g^2}
\label{eq:t_ss}
\end{equation}
with $\Delta_p = k-\omega_0$.

Comparing $t_{ss}$ in \eqref{eq:t_ss} with $t_{\Lambda}$ in \eqref{eq:t_lambda}, we see that the responses of the two systems are similar in the single-photon regime. The role of the control field $\Omega_c$ is now played by the dipole coupling $g$ and the decay rate $\Gamma_{20}$ is replaced by the decay rate $\gamma$ (when $\Gamma_{21} = 0$), up to constant factors. The stable state in the $\Lambda$ system, necessary for EIT to occur, is now provided by the disconnected atom. We plot $t_{ss}$ as a solid black line in \figurepanel{fig:results}{b} for $g = \gamma/8$, showing the similarity to $t_{\Lambda}$ in \figurepanel{fig:results}{a}.

Just like in a $\Lambda$ system, from the expression for the transmission coefficient we can identify the EIT regime (and likewise the ATS regime). The two complex poles of $t_{ss}$ are
\begin{equation}
    \mathcal{Z_{\pm}} = -\frac{i}{4} \mleft ( \gamma \pm \sqrt{\gamma^2 - 16 g^2} \mright ),
\end{equation}
which are purely imaginary for $ g < \gamma/4 $, defining the EIT regime. 

The comparison with a $\Lambda$ system breaks down outside the single-photon sector. By calculating the flux in \eqref{eq:flux} from the solution to the master equation in \eqref{eq:master}, for a weak coherent probe, we observe that the fluorescence is not quenched. This is illustrated  by the dashed line in \figurepanel{fig:results}{b}. Unlike in a $\Lambda$ system, the population is not trapped in a completely dark state in the steady state.


\subsubsection{Two separated small atoms coupled to the waveguide (``s-s'')}
\label{subsubsec:s-s}

We denote the separation of the two atoms, depicted in \figurepanel{fig:setup}{c}, as $\Delta x$, and position them one wavelength apart, $\Delta x = \lambda_0$, where their waveguide-mediated correlated decay is maximal. 
The wavelength $\lambda_0$ is determined by the resonant frequency of the first atom, $\lambda_0=2\pi/\omega_0$.
We note that this particular setup has attracted special interest recently due to its nonreciprocal property caused by inelastic scattering \cite{Fratini2014, Dai2015, Fratini2016, Mascarenhas2016, Fang2017, Muller2017, Hamann2018}.

To solve Eqs.~(\ref{eq:scattering1})-(\ref{eq:scattering3}),
we choose the excitation energies of the atoms as $\omega_1 = \omega_0$ and $ \omega_2 = \omega_0 + \delta$, where $\delta$ is a small detuning, and assume (i) the Markov approximation $k \Delta x \approx k_0 \Delta x$, where $k_0  = \omega_0$, and (ii) equal coupling strength $\gamma$ to the waveguide for each atom. This leads to the single-photon transmission coefficient
\bea
    t_{s-s} & = & 1 - \frac{i \gamma \mleft( \delta + 2\Delta_p \mright)}{2 \mleft(i \gamma + \delta + \Delta_p \mright) \Delta_p + i \gamma \delta} \nn\\ 
     & = & 1 - \frac{ 4 \gamma \Delta'_p}{4 \mleft( \gamma - i \Delta'_p \mright) \Delta'_p + i \delta^2},
\eea
where $\Delta_p' = \Delta_p + \delta/2$.
In \figurepanel{fig:results}{c}, we plot this transmission coefficient for and $\delta = \gamma/2$. We see that the transmission has an EIT-like peak when the drive is slightly off resonance with the first atom. 

The mapping to a $\Lambda$ system becomes even more clear by inspecting the poles of $t_{s-s}$,
\begin{equation}
    \mathcal{Z}_{\pm} = -\frac{\delta}{2} - \frac{i}{2} \mleft(\gamma \pm \sqrt{\gamma^2 - \delta^2} \mright).
\end{equation}
Note that the first term, $-\delta/2$, is real and shared by the two poles. Its effect is to shift the EIT peak from $\omega_0$ to $\omega_0 - \delta/2$, explaining the plot in \figurepanel{fig:results}{c}. The other terms in the two poles are purely imaginary for $\delta < \gamma$, which defines the EIT regime in this case. Thus, the role of the control field $\Omega_c$ is played by the detuning $\delta$, while $\Delta_p$ in the $\Lambda$ system is replaced by $\delta/2 + \Delta_p$. Unlike in the previous case with two dipole-coupled atoms (configuration ``ss''), here the mapping to a $\Lambda$ system also holds in the two-photon sector, where the fluorescence is fully quenched, $F = 0$ at the EIT frequency~\cite{Fang2017}; see the dashed line in \figurepanel{fig:results}{c}. 

\begin{figure}[t!]
    \centering
    \includegraphics[width=\columnwidth]{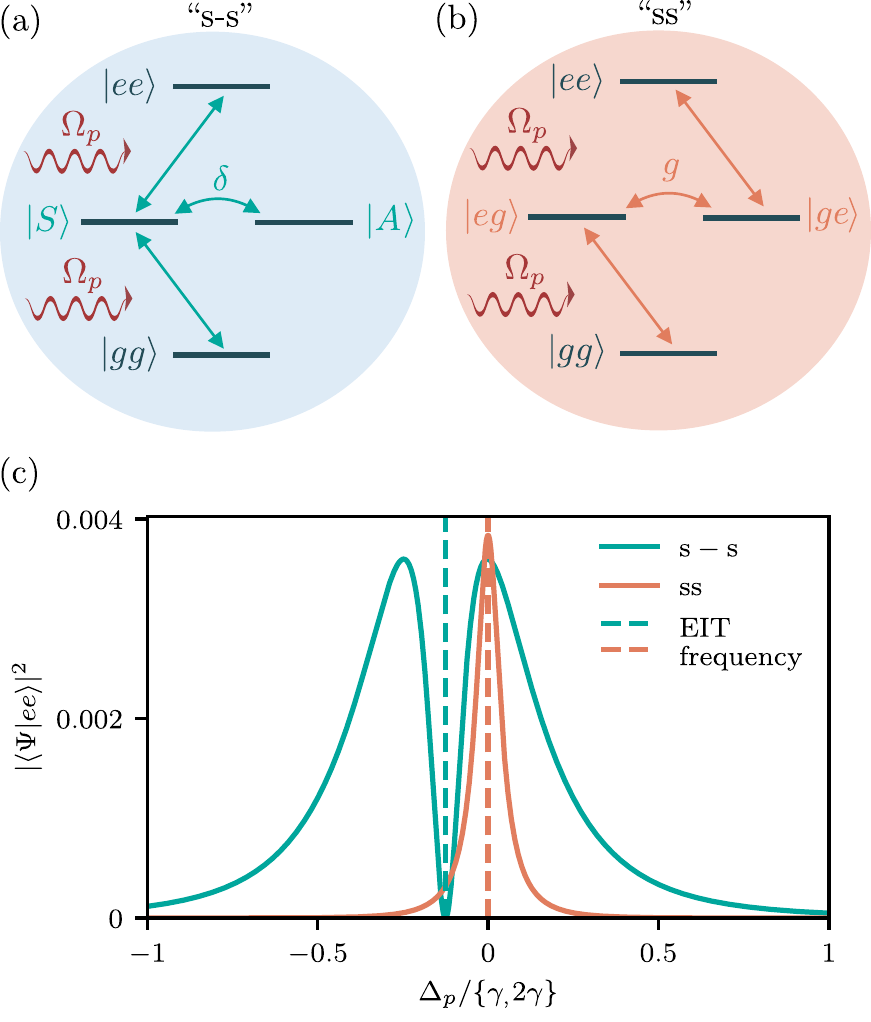}
    \caption{Energy diagrams explaining the presence or absence of a fluorescence quench in the ``s-s'' and ``ss'' setups. (a) Two detuned atoms, written in the symmetric/antisymmetric basis. In the steady state, the doubly excited state $\ket{ee}$ can never be excited since the transition from $\ket{A}$ to $\ket{ee}$ is forbidden. The system thus behaves like a $\Lambda$ system even outside of the single excitation regime. (b) Two dipole-coupled atoms. Because the transition from $\ket{ge}$ to $\ket{ee}$ is possible, the system can never be driven into a completely dark state, and the comparison with a $\Lambda$ system breaks down outside the single-excitation regime. (c) Excitation probability of the doubly excited state $\ket{ee}$ as a function of drive frequency for the ``s-s'' (turquoise line) and ``ss'' (orange line) systems in the steady state $\ket{\Psi}$. We used a coherent drive amplitude of $|\alpha|^2 = \gamma/5$ and $|\alpha|^2 = \gamma/10$ for ``s-s'' and ``ss'', respectively. Other parameters are the same as in Figs.~\figurepanelNoPrefix{fig:results}{c} and \figurepanelNoPrefix{fig:results}{b}, respectively.
    \label{fig:diagrams}}
\end{figure}

The reason why ``s-s'' has a quenched EIT peak while ``ss'' does not can be understood in terms of dressed states. By writing the ``s-s'' system Hamiltonian in the symmetric/antisymmetric basis with the coherent drive included, we have
\bea
H_E &=& \mleft( \Delta_p + \frac{\delta}{2} \mright) \mleft( \sigma^+_S \sigma^-_S + \sigma^+_A \sigma^-_A \mright) + \frac{\delta}{2} \mleft( \sigma^+_S \sigma^-_A + \sigma^+_A \sigma^-_S \mright) \nn\\
&& -i \frac{\Omega_p}{2} \mleft( \sigma^+_S - \sigma^-_S \mright),
\label{eq:H_E}
\eea
where $\sigma^{\pm}_S = \frac{1}{\sqrt{2}}(\sigma^{\pm}_1 + \sigma^{\pm}_2)$, and  $\sigma^{\pm}_A = \frac{1}{\sqrt{2}}(\sigma^{\pm}_1 - \sigma^{\pm}_2)$. It can be seen that only the symmetric state $\ket{S} = \sigma^+_S\ket{gg}$ couples to the waveguide probe field. By comparison with the above Hamiltonian for the $\Lambda$ system, \eqref{eq:Hamil_lambda}, we see the strong resemblance when $\Delta_c = 0$, since we can make the identifications $\ket{0} \leftrightarrow \ket{gg}$, $\ket{1} \leftrightarrow \ket{A}$, and $\ket{2} \leftrightarrow \ket{S}$. However, it would seem like the doubly excited state $\ket{ee}$ should make a crucial difference in the comparison with a $\Lambda$ system for both the ``s-s'' and ``ss'' setups. As shown in Figs.~\figurepanelNoPrefix{fig:diagrams}{a} and \figurepanelNoPrefix{fig:diagrams}{b}, both setups contain a coupling between one of the three lower levels and $\ket{ee}$.

The key turns out to be which of the lower energy levels couples to $\ket{ee}$. As shown in \figurepanel{fig:diagrams}{c}, $\ket{ee}$ is not occupied in the steady state when the ``s-s'' setup is driven at its EIT frequency, and this setup thus behaves like a proper $\Lambda$ system also outside the single-excitation regime. However, when driving the ``ss'' setup at its EIT frequency, the occupation probability for $\ket{ee}$ is nonzero, making this setup an effective $N$-type four-level system without any fluorescence quench.

In the ``s-s'' setup, the fact that it is the bright state $\ket{S}$ rather than the dark state $\ket{A}$ that couples to $\ket{ee}$ admits the existence of a \emph{pure dark steady state} for the system. Such a pure dark state must be an eigenstate of the Hamiltonian \textit{and} be annihilated by all jump operators in the master equation~\cite{Diehl2008, Kraus2008, Pichler2015}. The latter condition restricts us to the subspace spanned by $\ket{gg}$ and $\ket{A}$. Since the Hamiltonian $H_E$ in \eqref{eq:H_E} only couples these states to $\ket{S}$ at the EIT frequency $\Delta_p = - \delta/2$, it is possible to find a superposition of $\ket{gg}$ and $\ket{A}$ that is an eigenstate of $H_E$ with eigenvalue zero~\cite{Pichler2015}.
 
In the ``ss'' setup, it is instead the ``dark'' state $\ket{ge}$ that is coupled to $\ket{ee}$. Following the same reasoning as in the previous paragraph, a pure dark steady state in this setup must be a superposition of $\ket{gg}$ and $\ket{ge}$, both of which couple to $\ket{eg}$. But since $\ket{ge}$ also couples to $\ket{ee}$, a superposition of $\ket{gg}$ and $\ket{ge}$ cannot be an eigenstate of the Hamiltonian in \eqref{eq:H_E-ss}, and thus there is no pure dark steady state here. The consequence is that the ``ss'' setup does not behave as a $\Lambda$ system outside the single-excitation regime and therefore does not exhibit a fluorescence quench.

\subsection{Giant atoms}

Giant atoms were recently introduced as a way to tune the relaxation rates of individual artificial atoms \cite{Kockum2014, Kockum2020}. Additionally, when multiple giant atoms are placed close to each other in a waveguide, they can be arranged to create decoherence-free coherent interactions between the atoms~\cite{Kockum2018}. Here, we compare two different setups with giant atoms. The first setup, shown in \figurepanel{fig:setup}{d}, uses a single giant atom to create a dark state. A small atom is placed inside the giant atom in order to induce a waveguide-mediated coupling between the two atoms. We refer to this setup as ``Gs''. In the other giant-atom setup, we consider two giant atoms with braided coupling points. This arrangement of coupling points is the one that can create decoherence-free interactions~\cite{Kockum2018}, as demonstrated in a recent experiment~\cite{Kannan2020}, and we show that the same setup is capable of creating EIT-like phenomena. We refer to this setup as ``GG''; see \figurepanel{fig:setup}{e}. 


\subsubsection{One giant atom with one small atom inside (``Gs'')}
 
We set the separation between the two coupling points of the giant atom to $\Delta x = \lambda_0/2$ (with the first coupling point at $x = 0$),
such that the two relaxation pathways corresponding to emission at the two coupling points interfere destructively. The small atom is placed in-between the two giant-atom coupling points, positioned at the center $\Delta x/2$. The choice of position for the inner atom does not affect the analysis in any  way other than changing the effective coupling strength between the two atoms and also the location of the EIT peak. 

We solve Eqs.~(\ref{eq:scattering1})-(\ref{eq:scattering3}) using the Markovian approximation $k \Delta x \approx k_0\Delta x = \pi$, with the two atoms having equal excitation energy $\omega_0$, but unequal coupling strengths: $\gamma_G$ at each coupling point for the giant atom and $\gamma_s$ for the small atom. This gives the transmission coefficient
\begin{equation}
    t_{Gs} = 1 - \frac{ \gamma_s\Delta_p}{\mleft( \gamma_s - 2 i \Delta_p \mright) \Delta_p + 2 i \gamma_s\gamma_G}.
\end{equation}
We plot this transmission coefficent for $\gamma_G = \gamma_s/64$ in \figurepanel{fig:results}{d}. Once again, we observe an EIT-like transmission peak.

The poles of the transmission coefficient are located at
\begin{equation}
\mathcal{Z}_{\pm} = - \frac{i}{4} \mleft( \gamma_s \pm  \sqrt{ \gamma_s^2 - 16\gamma_G\gamma_s} \mright ),
\end{equation}
which are purely imaginary for $\gamma_G < \gamma_s/16$, defining the EIT regime. In this case, it is $\sqrt{2\gamma_G\gamma_s}$ which plays the role of the control field $\Omega_c$ in the mapping to a $\Lambda$ system. The dark giant atom behaves very similarly to the disconnected atom in the ``ss'' setup discussed in \secref{subsubsec:ss}, and no additional effects are introduced [compare Figs.~\figurepanelNoPrefix{fig:results}{b} and \figurepanelNoPrefix{fig:results}{d}]. The dipole coupling in the ``ss'' setup is replaced by a waveguide-mediated coherent coupling here, which is possible with a giant atom even though it does not decay into the waveguide~\cite{Kockum2018}. The comparison to the ``ss'' setup holds for both elastic and inelastic scattering properties: the ``Gs'' setup does not have any fluorescence quench accompanying the transmission peak.


\subsubsection{Two giant atoms in a braided configuration (``GG'')}

Until now, each system has been qualitatively different from the others in the way they achieve EIT-like peaks. However, with multiple giant atoms we have not found a different way to achieve EIT. Instead, motivated by the growing interest in giant atoms~\cite{Kockum2020}, and braided giant atoms specifically~\cite{Kockum2018, Kannan2020}, we simply show that this system is also capable of inducing EIT. 

We use the same approach to EIT as in the ``s-s'' setup (see \secref{subsubsec:s-s}), where the two small atoms are coupled through the waveguide and are slightly detuned from each other such that $\omega_1 = \omega_0$ and $\omega_2 = \omega_0 + \delta$. We set the distance between neighbouring coupling points of the giant atoms to $\Delta x/3$. We use the Markov approximation $k \Delta x/3 \approx k_0 \Delta x/3 = \pi$. We set the coupling strength at each connection point of both atoms to be $\gamma$. The transmission coefficient then becomes
\bea
    t_{GG} & = & 1 - \frac{2 i \gamma (\delta + 2 \Delta_p)}{(4 i \gamma + \delta + \Delta_p) \Delta_p + 2 i \gamma\delta} \nn\\
     & = & 1 - \frac{16 \gamma \Delta'_p}{4 \mleft( 4 \gamma - i \Delta'_p \mright) \Delta'_p + i \delta^2},
\eea
where $\Delta'_p = \Delta_p + \delta/2$.
We plot this transmission coefficient for $\delta = 2\gamma$ in \figurepanel{fig:results}{e}. We see that it exhibits an off-resonant transmission peak just like the ``s-s'' setup in \figurepanel{fig:results}{c}.

The poles of the transmission coefficient are located at
\begin{equation}
    Z_{\pm} = -\frac{\delta}{2} - \frac{i}{2}\mleft(4\gamma \pm \sqrt{16 \gamma^2 - \delta^2}\mright).
\end{equation}
From this, we see that the detuning introduces a shift in the EIT peak by $-\delta/2$ as in the ``s-s'' setup. The system is in the EIT regime for detunings $\delta < 4\gamma$. Also same as in the case of the ``s-s'' setup, the EIT-like transmission peak coincides with a fluorescence quench, as shown by the dashed line in \figurepanel{fig:results}{e}.


\section{Conclusion and outlook}
\label{sec:conclusion}

We have studied a number of setups in waveguide QED, where two atoms, either small or giant, exhibit narrow EIT-like transparency windows in certain parameter regimes without the presence of a strong external drive. We compared this behaviour to the EIT that can be achieved in a driven three-level $\Lambda$ system. To judge whether we truly were seeing analogs of EIT, we evaluated the existence of a fluorescence quench, i.e., cancellation of inelastic scattering, in all systems, and compared it to the fluorescence quench that accompanies EIT in a three-level $\Lambda$ system. 

To carry out these comparisons, we first performed a single-photon scattering calculation, where we derived a general set of equations that can be solved for the scattering coefficients in systems with an arbitrary number of atoms, both small and giant, with an arbitrary number of coupling points, at arbitrary positions, to a 1D waveguide. For the giant-atom coupling configurations, this means that the coupling points can be both ``braided'' and ``nested''~\cite{Kockum2018}, or surrounding a small atom. We used the scattering calculation to derive a criterion for being in the EIT regime, where the transparency is due to quantum interference, for each of the configurations depicted in \figref{fig:setup}. We then numerically solved a master equation in this parameter regime to calculate the inelastic photon flux, which allowed us to evaluate the existence of a fluorescence quench. The criteria for the EIT regime, and the existence of a fluorescence quench at the EIT frequency for each system, are summarized in \tabref{tab}.

All the systems we consider, shown in Figs.~\figurepanelNoPrefix{fig:setup}{b}-\figurepanelNoPrefix{fig:setup}{e}, consist of two two-level systems placed close to each other in a waveguide in some configuration. The difference between them is in the nature of the dark state involved in producing the narrow transparency windows seen in \figref{fig:results}. The key to these differences is highlighted in \figref{fig:diagrams}. Only when the atoms are detuned from each other, as in Figs.~\figurepanelNoPrefix{fig:setup}{c} and \figurepanelNoPrefix{fig:setup}{e}, can the two atoms be driven into a pure dark steady state by suppressing any occupation of the doubly excited state. The system then truly behaves like a $\Lambda$ system even beyond the single-excitation regime, which explains the fluorescence quench (i.e., lack of inelastic scattering). In the setups depicted in Figs.~\figurepanelNoPrefix{fig:setup}{b} and \figurepanelNoPrefix{fig:setup}{d}, there is no mechanism that prevents both atoms from being excited simultaneously by absorbing two photons,
which explains the lack of fluorescence quench observed in these systems.

Recently, several experiments have demonstrated synthesized EIT in waveguide QED~\cite{Koshino2013a, Inomata2014, Novikov2016, Long2018, Andersson2020, Vadiraj2020}. This was done either by embedding atoms in cavities or resonators to form an effective three-level system in terms of dressed states, or by engineering the decay rates of individual energy levels directly. Ultimately, these demonstrations relied on the applications of multiple coherent fields. We believe that two-levels systems offer an easier way to synthesize EIT, with potentially better performance due to its simplicity. All setups we have studied here are straightforward to implement in waveguide QED with superconducting circuits, and some may also be within reach for other implementations of waveguide QED. Our work can help direct future studies towards genuine EIT-like physics, and avoid confusion regarding the necessary ingredients for EIT to occur. Given the many applications of EIT and other phenomena in $\Lambda$ systems~\cite{Lukin2003, Fleischhauer2005}, the results presented here suggest that some of these applications can be transferred to the domain of waveguide QED without requiring a strong control field.

For example, in Ref.~\cite{Leung2012}, Leung et al.~studied a waveguide-QED-based quantum memory consisting of a 1D array of $N$ three-level atoms, analogous to its atomic-gas counterpart routinely used in laboratories nowadays. Our work indicates the possibility of replacing the three-level atoms with simpler two-level ones, possibly giant, which could greatly reduce the experimental overhead of controlling $N$ atoms via $N$ microwave transmission lines. It is worth noting, from a theoretical perspective, that the analysis in Ref.~\cite{Leung2012} involved a transfer-matrix approach to calculate the transmission coefficient $t$; however, this method is not straightforward to generalize if braided/nested giant atoms are present due to the feedback loops from the atoms, so to investigate the single-photon regime by solving Eqs.~(\ref{eq:scattering1})-(\ref{eq:scattering3}) is likely the most rigorous (though tedious) approach.

Another example is the recent studies on the ``s-s''-setup-based passive quantum rectifier~\cite{Fratini2014, Dai2015, Fratini2016, Mascarenhas2016, Fang2017, Muller2017, Hamann2018}, in which the effect is maximized when the dark state is driven resonantly~\cite{Dai2015, Fang2017}. Based on the similarity between the ``s-s'' and ``GG'' setups we pointed out above and the high flexibility of giant-atom configurations~\cite{Kockum2014, Kockum2018}, it is possible that a configuration of giant atoms [such as a more complex braiding than the one considered in \figurepanel{fig:setup}{e}] could be engineered to achieve a similar or even stronger nonreciprocal effect. We leave investigations of this possibility for future work.


\begin{acknowledgments}

AA acknowledges support from the Swedish Research Council (grant number 2017-04197).
YLLF is supported in part by BNL LDRD 19-002 and by New York State Urban Development Corporation, d/b/a Empire State Development, under contract No.\ AA289.
AFK acknowledges support from the Swedish Research Council (grant number 2019-03696), and from the Knut and Alice Wallenberg Foundation through the Wallenberg Centre for Quantum Technology (WACQT).

\end{acknowledgments}


\bibliography{apssamp}

\end{document}